\documentclass[11pt]{article}

\begin{document}

\date{\empty}

\title{\textbf{On the stability of static ghost cosmologies}}
\author{John D. Barrow$^1$ and Christos G. Tsagas$^2$\\ {\small $^1$~DAMTP, Centre for Mathematical Sciences}\\ {\small University of Cambridge, Wilberforce Road, Cambridge CB3 0WA, UK}\\ {\small $^2$~Section of Astrophysics, Astronomy and Mechanics, Department of Physics}\\ {\small Aristotle University of Thessaloniki, Thessaloniki 54124, Greece}}

\maketitle

\begin{abstract}
We consider the classical linear stability of a static universe filled with a non-interacting mixture of isotropic radiation and a ghost scalar field. Unlike the conventional Einstein static model, this cosmology is stable against homogeneous and isotropic perturbations. This is shown by means of exact oscillatory solutions about the original static state. We also examine the linear response of the static ghost universe to all types of inhomogeneous fluctuations, namely density, vorticity and gravitational-wave perturbations. The results show that the static background constrains the linear evolution of these distortions, to the extent that density perturbations remain time invariant, vortical distortions vanish and gravitational waves oscillate with constant amplitude. We discuss the potential implications of these results for past-eternal initial states in classical general relativistic cosmology.
\end{abstract}

\section{Introduction}
Cosmologists have retained an interest in the consequences of admitting `ghost' scalar fields into the Universe. These fields may have a negative energy density~\cite{G1}, and are far from new, having been incorporated in the early study of steady-state cosmology~\cite{H,Mc}, in the investigation of scalar fields~\cite{BB}, phantom dark matter~\cite{C,BFK}, and $\kappa$-essence~\cite{A-PMS}. Although ghost-fields may be unstable at the quantum level~\cite{CHT,CJM}, they provide a simple theoretical laboratory for exploring the physical consequences of cyclic closed universes which bounce at a finite radius. In particular, Barrow et al~\cite{BKM} examined the unusual consequences that arise for theories of varying 'constants' in such oscillating universes. Ellis and collaborators \cite{EM,EMT} have also considered cosmologies that emerge into expansion from a past eternal Einstein static state, reminiscent of the original Eddington-Lema\^{\i}tre cosmology favoured by Eddington as an infinitely old universe that can nonetheless be thermodynamically young~\cite{E,BM}. One further consequence of these ghost cosmologies is the existence of static solutions with new properties. In this paper we will examine the stability of Einstein static universes in these theories. Two new features will be of interest. First, the existence of almost static solutions that oscillate about a finite radius; and second, the classical linear stability of certain ghost-like static models under all three types of inhomogeneous perturbations (i.e.~scalar, vector and tensor modes). The inhomogeneous situation shares some analogies with the stability shown by the Einstein static universe in the presence of conventional (non-ghost) matter fields, as discussed in~\cite{Ha}-\cite{BEMT}.\footnote{The stability analysis discussed there is also rendered delicate because the Einstein static universe has compact space sections and Killing vectors and so is a conical point in the space of all solutions to the Einstein equations which exhibit linearisation instability about it unless higher-order constraints are imposed~\cite{BD}-\cite{LU}.}

Throughout this paper we assume a pseudo-Riemannian spacetime with a Lorentzian metric of signature ($-\,+\,+\,+$). We also adopt the 1+3 covariant approach to general relativity and cosmology, referring the reader to older and more recent reviews of the formalism for further reading and details~\cite{El}-\cite{TCM}.

\section{Ghost scalar fields}\label{sGSF} 
At the basis of the 1+3 covariant formalism is the concept of the
fundamental observers, which when introduced allow for a unique `threading' of the spacetime into time and 3-dimensional space. The worldlines of these observers are tangent to a timelike 4-velocity field, relative to which physical quantities and equations decompose into their timelike and spacelike components. The formalism combines mathematical compactness with physical transparency and has been used to a variety of cosmological are general relativistic studies. Here we will use the covariant approach to study cosmological models containing a ghost scalar field ($\psi$) with zero potential (i.e.~$V(\psi)=0$) and a Lagrangian of the form~\cite{BKM}
\begin{equation}
\mathcal{L}_{\psi}= {\frac{1}{2}}\,\nabla_a\psi\nabla^a\psi\sqrt{-g}\,,
\label{Lpsi}
\end{equation}
where $g$ is the determinant of the spacetime metric ($g_{ab}$) and $\nabla_a $ is the standard covariant derivative operator. Then, the associated stress-energy tensor reads
\begin{equation}
T^{(\psi)}_{ab}= -\nabla_a\psi\nabla_b\psi+ {\frac{1}{2}}\,\nabla_c\psi\nabla^c\psi g_{ab}\,.  \label{Tpsi1}
\end{equation}

To achieve an 1+3 decomposition of the $\psi$-field, we assume that $\nabla_{a}\psi$ is timelike (namely that $\nabla_{a}\psi\nabla^{a}\psi<0$) over a given spacetime region. In that case $\nabla_{a}\psi$ is normal to the spacelike hypersurfaces $\psi(x_{a})=$~constant and defines our fundamental timelike 4-velocity field
\begin{equation}
u_{a}= -{\frac{1}{\dot{\psi}}}\,\nabla_{a}\psi\,,  \label{ua}
\end{equation}
with $\dot{\psi}=u^{a}\nabla_{a}\psi\neq0$. This guarantees that $\dot{\psi}^{2}=-\nabla_{a}\psi\nabla^{a}\psi>0$ and that $u_{a}u^{a}=-1$ as required. The above 4-velocity also defines the fundamental time direction and the tensor $h_{ab}=g_{ab}+u_{a}u_{b}$ that projects into the 3-space orthogonal to $u_{a}$. The projection tensor also provides the covariant derivative $\mathrm{D}_a=h_a{}^b\nabla_b$, which operates in the observers' instantaneous rest space.

In covariant terms, the kinematics of the $u_a$-congruence (\ref{ua}) are determined by means of the irreducible decomposition~\cite{E,TCM}
\begin{equation}
\nabla_bu_a= {\frac{1}{3}}\,\Theta h_{ab}+ \sigma_{ab}+ \omega_{ab}- A_au_b\,,  \label{Nbua}
\end{equation}
where $\Theta$, $\sigma_{ab}$ and $\omega_{ab}$ respectively describe the average volume expansion/contraction, shear distortions and the vorticity of the flow. Following our 4-velocity choice, the later vanishes identically (i.e.~$\omega_{ab}=0$), making the $u_a$-field irrotational.\footnote{An direct consequence of our 4-velocity choice (see (\ref{ua})) is that $\mathrm{D}_a\psi=0$. This in turn guarantees that $\omega_{ab}=\mathrm{D}_{[b}u_{a]}=0$ and the irrotational nature of the flow (e.g.~see~\cite{TCM}).} The 4-acceleration vector ($A_a$) describes non-gravitational/non-inertial forces and is given by
\begin{equation}
A_a= -{\frac{1}{\dot{\psi}}}\,\mathrm{D}_a\dot{\psi}\,.  \label{Aa}
\end{equation}
The above, which may be seen as the momentum conservation law of the $\psi$-field (compare to Eq.~(\ref{lsystem1}b) in \S~\ref{ssLDI}), implies that $\dot{\psi}$ acts as an acceleration potential.

Introducing the $u_{a}$-frame (\ref{ua}) also facilitates a convenient fluid-like description of the $\psi$-field. In particular, with respect to the fundamental observers, the energy-momentum tensor (\ref{Tpsi1}) recasts into
\begin{equation}
T_{ab}^{(\psi)}= \rho^{(\psi)}u_{a}u_{b}+ p^{(\psi)}h_{ab}\,,  \label{Tpsi2}
\end{equation}
where
\begin{equation}
\rho ^{(\psi)}= p^{(\psi)}= -{\frac{1}{2}}\,\dot{\psi}^{2}\,.  \label{psiES}
\end{equation}
The right-hand side of (\ref{psiES}) is negative by default, ensuring that $\rho^{(\psi)}<0$ and consequently the ghost nature of the $\psi$-field.

In the presence of a non-vanishing potential, scalar fields do not generally behave like barotropic fluids. Here, however, the potential has been set to zero and a barotropic description of the $\psi$-field is therefore possible. According to (\ref{psiES}), our scalar field corresponds to a `stiff' medium with effective barotropic index $w^{(\psi)}=p^{(\psi)}/\rho^{(\psi)}=1$ and a corresponding sound-speed given by $c_{\psi}^2=\mathrm{d}p^{(\psi)}/\mathrm{d}\rho^{(\psi)}=1$.

\section{Homogeneous ghost cosmologies}\label{sHGC} 
\subsection{Static spatially closed models}\label{ssSSCM} 
The evolution of a Friedmann-Robertson-Walker (FRW) cosmology with nonzero spatial curvature and a non-vanishing cosmological constant is monitored by the set
\begin{equation}
H^2= {\frac{1}{3}}\,\left(\rho+\Lambda\right)- {\frac{K}{a^2}}\,, \hspace{15mm} \dot{H}= -H^2- {\frac{1}{6}}\,(\rho+3p)+ {\frac{1}{3}}\Lambda  \label{FR}
\end{equation}
and
\begin{equation}
\dot{\rho}= -3H(\rho+p)\,.  \label{ceqn}
\end{equation}
In the above $\rho$ and $p$ represent the total energy density and pressure of the matter, $K=0,\pm1$ is the 3-curvature index and $\Lambda$ is the cosmological constant. In addition $H=\dot{a}/a$ is the Hubble parameter of the model and $a$ is the cosmological scale factor. To close this system one needs to supply an (effective) equation of state for the (total) matter sources.

Consider a static cosmology with zero cosmological constant and positive spatial curvature. Then, the cosmological scale factor is time-independent ($a=a_{0}=$~constant) and can be seen as the radius of the universe. Setting $\Lambda=0$, $K=+1$ and $H=0=\dot{H}$, Eqs.~(\ref{FR})-(\ref{ceqn}) recast into the constrains
\begin{equation}
\rho= {\frac{3}{a_{0}^{2}}}\,, \hspace{15mm} \rho+ 3p= 0 \hspace{15mm} \mathrm{and} \hspace{15mm} \dot{\rho}=0\,,  \label{static}
\end{equation}
respectively.\footnote{A positive 3-curvature allows us to relate more directly with the recent Emergent-universe scenario~\cite{EM,EMT}, as well as the much earlier Einstein static and the Eddington-Lema\^{i}tre models~\cite{E}-\cite{BEMT}. A nonzero cosmological constant is not necessary for our purposes, since its role is now played by the dynamical, ghost-like scalar field.} The former of the above shows that the size of our model is inversely proportional to the energy density of its matter content, namely that $a_{0}=\sqrt{3/\rho}$. The second guarantees that the effective barotropic index of the system is $w=p/\rho=-1/3$.

Suppose now that matter forms non-interacting mixture of isotropic radiation, with $p^{(\gamma)}=\rho^{(\gamma)}/3$, supplemented by a ghost scalar field ($\psi$) with zero potential and an effective equation of state given by (\ref{psiES}). When both components are comoving, their total energy density and pressure, relative to a fundamental observer, are $\rho=\rho^{(\gamma)}+\rho^{(\psi)}$ and $p=p^{(\gamma)}+p^{(\psi)}$ respectively. In that case expressions (\ref{static}) reduce to
\begin{equation}
\rho^{(\gamma)}+ \rho^{(\psi)}= {\frac{3}{a^{2}}}\,, \hspace{25mm}
\rho^{(\gamma)}+ 2\rho^{(\psi)}= 0  \label{con12}
\end{equation}
and
\begin{equation}
\dot{\rho}^{(\gamma)}+ \dot{\rho}^{(\psi)}= 0\,.  \label{con3}
\end{equation}
Of the above constraints, (\ref{con12}a) combines with (\ref{con12}b) to ensure that the energy densities of the two species are directly related, according to
\begin{equation}
\rho^{(\psi)}= -{\frac{1}{2}}\,\rho^{(\gamma)}= -\rho\,.  \label{rhos}
\end{equation}
This result then, together with (\ref{con12}a), leads to
\begin{equation}
\rho^{(\gamma)}= {\frac{6}{a_{0}^{2}}}\Leftrightarrow a_{0}= \sqrt{{\frac{6}{\rho^{(\gamma)}}}} \hspace{15mm} \mathrm{and} \hspace{15mm} \rho^{(\psi)}= -{\frac{3}{a_{0}^{2}}}\Leftrightarrow a_{0}= \sqrt{-{\frac{3}{\rho^{(\psi)}}}}\,,  \label{con4}
\end{equation}
expressing the radius of our static model in terms of the energy densities of the individual matter components. We finally note that, for non-interacting fluids, the conservation law (\ref{con3}) is also separately satisfied (i.e.~$\dot{\rho}^{(\gamma)}=0= \dot{\rho}^{(\psi)}$).

\subsection{Almost-static oscillatory models}\label{ssASOM} 
Before looking into inhomogeneous cosmologies with ghost-like scalar fields, we will provide an example of a soluble almost-static homogeneous model. The Friedmann equation for the scale factor, $a(t)$, of a closed universe containing a ghost scalar field with density $\rho^{(\psi)}=-\dot{\psi}^{2}/2=-\Psi a^{-6}<0,$ and radiation with density $\rho^{(\gamma)}=\Gamma a^{-4},$ reads
\begin{equation}
\frac{\dot{a}^{2}}{a^{2}}= -\frac{\Psi}{a^{6}}+ \frac{\Gamma}{a^{4}}- \frac{1}{a^{2}},  \label{frw}
\end{equation}
where $\Psi$ and $\Gamma$ are positive constants. Expressed in terms of conformal time ($\eta$, with $\mathrm{d}\eta=a^{-1}\mathrm{d}t$), the above integrates to give
\begin{equation}
a^{2}(\eta)= \frac{1}{2}\left\{\Gamma+\sqrt{\Gamma^{2}-4\Psi} \sin\left[2(\eta+\eta_{0})\right]\right\}\,,  \label{eta}
\end{equation}
with $\eta_{0}$ constant when $\Gamma^{2}\geq4\Psi$.

Identifying the global maximum and the global minimum of the expansion with $a^2_{\max}=[\Gamma+\sqrt{\Gamma^{2}-4\Psi}]/2$ and $a^2_{\min}=[\Gamma-\sqrt{\Gamma^{2}-4\Psi}]/2$ respectively, we find that the scale factor evolution is given by
\begin{equation}
a^2(\eta)= \frac{1}{2} \left\{a_{\max}^{2}+a_{\min}^{2} +(a_{\max}^{2}-a_{\min}^{2}) \sin\left[2(\eta+\eta_{0})\right]\right\}\,.  \label{sol2}
\end{equation}
Given that both $a_{\max}$ and $a_{\min}$ are finite (recall that $\Gamma$, $\Psi\neq0$), this solution oscillates about an also finite radius between successive maxima and minima.\footnote{For mathematical simplicity we have specialised the matter content of the model to be radiation plus the ghost field. The qualitative behavior of the Friedmann equation would be similar if the radiation was replaced by any perfect fluid with $\rho>p>-\rho/3.$}. If we had chosen $K=0$ then there would have been a single minimum (and no maximum) with collapse for $a<a_{\min}$ and expansion for $a>a_{\min} $. Since $\dot{\psi}=\sqrt{2\Psi}a^{-3},$ the scalar field evolution given by~\cite{BKM}
\begin{equation}
\psi= \pm2\tan^{-1}\left[\frac{\Gamma\tan \left(\eta+\eta_{0}\right) +\sqrt{\Gamma^{2}-4\Psi}}{2\sqrt{\Psi}}\right] .
\end{equation}
With $a_{\max}\gg a_{\min}$, we have $a_{\min}=\sqrt{\Psi/\Gamma}$ and $a_{\max}=\sqrt{\Gamma}$. We can then see that the bounce duration is $\Delta t\sim a_{\min}^{2}/a_{\max}$. Since $\dot{\psi}\sim\sqrt{6}\Gamma^{3/2}/\Psi$ near the bounce, we find that $\Delta\psi\sim\sqrt{6}$, independently of initial conditions, during each bounce.

The extreme case $\Gamma^{2}=4\Psi$ is a static universe with $a^2=a_{\max}^2=a_{\min}^2=\Gamma/2=\sqrt{\Psi}$. Setting $\dot{a}=0$ and $\ddot{a}=0$, we can see that this case is realized when $\rho_{\psi}= -{\tilde{\rho}}_{r}/2$, giving $a=\sqrt{6/{\tilde{\rho}_{r}}}$. We can see explicitly that the solution $a^2=\Gamma/2$ is stable against homogeneous and isotropic conformal perturbations that do not change the curvature of the model. Indeed, deviations from the static model are monitored by the parameter
\begin{equation}
\epsilon= \sqrt{\Gamma^2-4\Psi}= a_{\max}^2-a_{\min}^2\,.  \label{epsilon}
\end{equation}
Then, following (\ref{eta}) and (\ref{sol2}), the perturbed spacetime oscillates about the static solution with (arbitrary) amplitude determined by $\epsilon$. In particular, expressed in terms of the above defined parameter, solutions (\ref{eta}), (\ref{sol2}) assume the perturbative form
\begin{equation}
a^2(\eta)= \frac{1}{2} \left\{\Gamma+ \epsilon\sin\left[2(\eta+\eta_{0})\right]\right\}\,,  \label{sol3}
\end{equation}
which describes an almost static oscillatory (or cyclic) universe of constant amplitude. This result is in sharp contrast to the instability shown by the conventional Einstein-static cosmology~\cite{Ha}-\cite{BEMT}.

We will now turn to consider the linear stability of these static solutions against all three types of inhomogeneous perturbations namely scalar, vector and pure-tensor (i.e.~gravitational-wave) distortions. In the latter case we will also take a step into the nonlinear regime, by looking at the static model's response to gravity-wave anisotropies of arbitrary magnidude.

\section{Perturbed static ghost cosmologies}\label{sPSGC} 
\subsection{Describing inhomogeneity}\label{ssDI} 
In accord with the 1+3 covariant approach to cosmology, spatial inhomogeneities in the density distribution of the matter fields are described via the associated comoving fractional density gradients. For the total fluid, these are given by the dimensionless ratio~\cite{EB}
\begin{equation}
\Delta _{a}={\frac{a}{\rho }}\,\mathrm{D}_{a}\rho \,,  \label{Delta1}
\end{equation}
with $\mathrm{D}_{a}=h_{a}{}^{b}\nabla _{b}$ representing the covariant derivative operator in the observers rest space. Similarly, the gradients
\begin{equation}
\Delta_{a}^{(\gamma)}= {\frac{a}{\rho^{(\gamma)}}}\, \mathrm{D}_{a}\rho^{(\gamma)} \hspace{15mm} \mathrm{and} \hspace{15mm} \Delta_{a}^{(\psi)}= {\frac{a}{\rho^{(\psi)}}}\, \mathrm{D}_{a}\rho^{(\psi)}\,,  \label{Delta23}
\end{equation}
monitor inhomogeneities in the densities of the two individual matter components. When the background model is spatially homogeneous all of the above gradients vanish and therefore describe linear density perturbations in a gauge invariant manner~\cite{SW}. Also, definitions (\ref{Delta1}) and (\ref{Delta23}) imply that
\begin{equation}
\Delta_{a}= {\frac{1}{\rho}} \left(\rho^{(\gamma)}\Delta_{a}^{(\gamma)}+ \rho^{(\psi)}\Delta_{a}^{(\psi)}\right)\,.  \label{Delta2}
\end{equation}
The density gradients are supplemented by additional inhomogeneity variables, of which the most important describes fluctuations in the volume expansion ($\Theta$). The latter is represented by the orthogonally projected gradient
\begin{equation}
\mathcal{Z}_{a}= a\mathrm{D}_{a}\Theta\,,  \label{cZa}
\end{equation}
which is also independent of the gauge choice.

Consider the background model described in \S~\ref{ssSSCM} and perturb it by introducing weak inhomogeneities and anisotropies. To linear order, definitions (\ref{Delta1}), (\ref{Delta23}) and (\ref{cZa}) read
\begin{equation}
\Delta _{a}= {\frac{a_{0}}{\rho}}\, \mathrm{D}_{a}\rho\,, \hspace{10mm} \Delta_{a}^{(\gamma)}= {\frac{a_{0}}{\rho^{(\gamma)}}}\, \mathrm{D}_{a}\rho^{(\gamma)}\,, \hspace{10mm} \Delta_{a}^{(\psi)}= {\frac{a_{0}}{\rho^{(\psi)}}}\,\mathrm{D}_{a}\rho^{(\psi)}  \label{lDelta}
\end{equation}
and
\begin{equation}
\mathcal{Z}_{a}= a_{0}\mathrm{D}_{a}\Theta\,,  \label{lcZa}
\end{equation}
respectively. Also to first order, relation (\ref{Delta2}) reduces to
\begin{equation}
\Delta_{a}= 2\Delta_{a}^{(\gamma)}- \Delta_{a}^{(\psi)}\,,  \label{Delta3}
\end{equation}
since $\rho^{(\psi)}=-\rho^{(\gamma)}/2=-\rho$ in the background (see conditions (\ref{con12})).

\subsection{Linear density inhomogeneities}\label{ssLDI} 
In non-interacting multi-component systems, the individual members observe separate conservation laws. Since the $\psi$-field does not interact with the radiative component, their associated linear momentum conservation formulae are
\begin{equation}
a_0\left(1+w^{(i)}\right)A_a= -c_s^{2(i)}\Delta_{a}^{(i)}\,,  \label{limcl}
\end{equation}
with $i=\gamma,\,\psi$ and $A_a$ being the common 4-acceleration vector (e.g.~see~\cite{TCM}). In the absence of interactions, linear density inhomogeneities in each one of the species evolve according to
\begin{equation}
\dot{\Delta}_{a}^{(i)}= -\left(1+w^{(i)}\right)\mathcal{Z}_{a}\,,
\label{ldotDelta}
\end{equation}
where
\begin{equation}
\dot{\mathcal{Z}}_{a}= -{\frac{1}{2}}\,\rho\left(1+3c_{s}^{2}\right)
\Delta_{a}- {\frac{a_{0}}{2}}\,\rho\,(1+3w)A_{a}+ a_{0}\mathrm{D}_{a}\mathrm{D}^{b}A_{b}\,,  \label{ldotcZ}
\end{equation}
to linear order. Note that $\rho$, $p$, $w$ and $c_{s}^{2}$ correspond to the total fluid, and $\Delta_{a}$ is given by (\ref{Delta1}).

When dealing with a perturbed static universe filled with a non-interacting mixture of radiation and a ghost scalar field, we have $w^{(\gamma)}=1/3=c_s^{2(\gamma)}$ and $w^{(\psi)}=1=c_s^{2(\psi)}$ at the zero perturbative level. Hence, the linear expressions (\ref{ldotDelta})-({\ref{limcl}) lead to
\begin{equation}
4a_0A_a= -\Delta_{a}^{(\gamma)}\,, \hspace{15mm} 2a_0A_a=
-\Delta_{a}^{(\psi)}  \label{lsystem1}
\end{equation}
and
\begin{equation}
\dot{\Delta}_{a}^{(\gamma)}= -{\frac{4}{3}}\,\mathcal{Z}_{a}\,, \hspace{15mm}
\dot{\Delta}_{a}^{(\psi)}= -2\mathcal{Z}_{a}\,.  \label{lsystem2}
\end{equation}
The first of these sets implies that $\Delta_a^{(\gamma)}=2\Delta_a^{(\psi)}$, while second ensures that $\Delta_a^{(\gamma)}=2\Delta_a^{(\psi)}/3+\mathcal{C}$ (with $\mathcal{C}$ being time-independent). Therefore, to
linear order
\begin{equation}
\Delta_a^{(\gamma)}= 2\Delta_a^{(\psi)}= {\frac{3}{2}}\,\mathcal{C}\,.
\label{lDeltas}
\end{equation}
In other words, linear density inhomogeneities (both in the individual species and the total fluid -- see Eq.~(\ref{Delta3})) remain constant in time. This result seems to reflect the high symmetry of the unperturbed model and the properties of the two matter fields. In particular, the static background, the absence of a cosmological constant and the non-interacting nature of the material mixture mean that the energy densities of the constituent species are directly related (see constraints~(\ref{con12})-(\ref{con4})). When the two matter fields also share the same 4-velocity, as it happens here, the aforementioned constraints lead to the momentum conservation laws (\ref{lsystem1}).\footnote{In general, one could assign different 4-velocities to the two fluids, in which case relative-motion terms will also appear in the linear equations (e.g.~see \S~2.4 and \S~3.3 in~\cite{TCM}).} The latter, together with the propagation formulae (\ref{lsystem2}), guarantee that any linear density inhomogeneities that might exist will not change in time.\footnote{The time independence of $\Delta_a^{(\gamma)}$ and $\Delta_a^{(\psi)}$ implies that the linear expansion inhomogeneities must vanish at all times (see Eqs.~(\ref{lsystem2})). This imposes a strict condition to the right-hand side of (\ref{ldotcZ}), which must remain zero at all times. Checking the consistency of the linear condition goes beyond the scope of this article.}

Result (\ref{lDeltas}) also guarantees the time independence of linear density perturbations (described via the scalar $\Delta=a_0\mathrm{D}^a\Delta_a$) as well as of shape distortions (monitored by means of the trace-free tensor $\Delta_{\langle
ab\rangle}=a_0\mathrm{D}_{\langle b}\Delta_{a\rangle}$). On the other hand, vortical perturbation in the density distribution of the species (described via the antisymmetric tensor $\Delta_{[ab]}=a_0\mathrm{D}_{[b}\Delta_{a]}$) vanish identically. To verify this recall that (e.g.~see Appendix A.3 in~\cite{TCM})
\begin{equation}
\mathrm{D}_{[b}\mathrm{D}_{a]}\rho= \dot{\rho}\,\omega_{ab}= 0\,,  \label{Del[ab]}
\end{equation}
with the null result guaranteed on all perturbative levels by the irrotational nature of the $u_a$-frame (see \S~\ref{sGSF} earlier).

\subsection{Linear gravitational-wave perturbations}\label{ssLGWP}
So far, our analysis has shown that scalar (density) perturbations are time independent, while vector (rotational) distortions vanish. The remaining third type, concerns pure tensor fluctuations. Gravitational waves are covariantly described by the transverse parts of the electric ($E_{ab}$) and the magnetic ($H_{ab}$) components of the Weyl tensor. To linear order, these are isolated by switching off both the scalar and the vector perturbations (to guarantee that $\mathrm{D}^bE_{ab}=0=\mathrm{D}^bH_{ab}$ -- e.g.~see~\cite{TCM}). On a static, FRW-type background, the linear
propagation of these two fields is monitored by the set
\begin{equation}
\dot{E}_{ab}= -{\frac{1}{2}}\,\rho(1+w)\sigma_{ab}+ \mathrm{curl}H_{ab}
\label{dotE}
\end{equation}
and
\begin{equation}
\dot{H}_{ab}= -\mathrm{curl}E_{ab}\,.  \label{dotH}
\end{equation}
where $\mathrm{curl}E_{ab}=\varepsilon_{cd\langle a} \mathrm{D}^cE_{b\rangle}{}^d$ for any traceless and transverse tensor $E_{ab}$. Also to first order, the magnetic part of the Weyl field is related to the shear by means of
\begin{equation}
\mathrm{curl}H_{ab}= {\frac{3}{a_0^2}}\,\sigma_{ab}- \mathrm{D}^2\sigma_{ab}\,,  \label{Hab}
\end{equation}
with $\mathrm{D}^2=\mathrm{D}^a\mathrm{D}_a$ representing the orthogonally projected covariant Laplacian and $a_0$ the radius of the background 3-D hypersurfaces. The above ensures the direct interconnection between the magnetic Weyl field and the shear at the linear perturbative level. This in turn allows us to replace Eq.~(\ref{dotH}) with the shear propagation formula
\begin{equation}
\dot{\sigma}_{ab}= -E_{ab}\,.  \label{dotsigma}
\end{equation}
The latter combines with (\ref{dotE}) to give the wave-equation
\begin{equation}
\ddot{\sigma}_{ab}= {\frac{1}{2}}\,\rho(1+w)\sigma_{ab}- {\frac{3}{a_0^2}}\,\sigma_{ab}+ \mathrm{D}^2\sigma_{ab}\,,  \label{ddotsigma1}
\end{equation}
of the transverse part of the shear tensor. When the static background contains a mixture of non-interacting radiation and a ghost scalar field, $w=-1/3$ and $\rho=3/a_0^2$ (see \S~\ref{ssSSCM}). Applied to this environment, Eq.~(\ref{ddotsigma1}) reduces to
\begin{equation}
\ddot{\sigma}_{ab}= -{\frac{2}{a_0^2}}\,\sigma_{ab}+ \mathrm{D}^2\sigma_{ab}\,.  \label{ddotsigma2}
\end{equation}

Our next step is to Fourier decompose the last expression. Introducing the standard tensor harmonic functions, we may write $\sigma_{ab}=\sum_{n}\sigma_{(n)}\mathcal{Q}_{ab}^{(n)}$, where $\mathcal{Q}_{ab}^{(n)}$ are the aforementioned harmonics with $\mathcal{Q}_{ab}^{(n)}=\mathcal{Q}_{\langle ab\rangle}^{(n)}$, $\dot{\mathcal{Q}}_{ab}^{(n)}=0=\mathrm{D}^{b}\mathcal{Q}_{ab}^{(n)}$ and\footnote{Angled brackets denote the symmetric and trace-free part of orthogonally projected second rank tensors.}
\begin{equation}
\mathrm{D}^{2}\mathcal{Q}_{ab}^{(n)}= -\left({\frac{n}{a_{0}}}\right)^{2}\mathcal{Q}_{ab}^{(n)}\,.  \label{LB}
\end{equation}
Substituting decomposition $\sigma_{ab}=\sum_{n}\sigma_{(n)}\mathcal{Q}_{ab}^{(n)}$ into Eq.~(\ref{ddotsigma2}), the harmonics decouple and we arrive at the wave equation of the $n$-th shear mode
\begin{equation}
\ddot{\sigma}_{(n)}= -\left({\frac{n^{2}+2}{a_{0}^{2}}}\right)\,\sigma_{(n)}\,,  \label{ddotsigma3}
\end{equation}
where $n^{2}\geq3$ due to the compactness of the unperturbed spatial sections. Expression (\ref{ddotsigma3}) governs the linear evolution of gravitational waves on a static, FRW-type background that has spherical spatial geometry and contains a mixture of non-interacting black-body radiation with a ghost scalar field. The solution of (\ref{ddotsigma3}) has the simple oscillatory form
\begin{equation}
\sigma _{(n)}=\mathcal{C}_{1} \sin\left[\left({\frac{\sqrt{n^{2}+2}}{a_{0}^{2}}}\right)t\right] +\mathcal{C}_{2} \cos\left[\left({\frac{\sqrt{n^{2}+2}}{a_{0}^{2}}}\right)t\right]\,,  \label{sigma}
\end{equation}
which ensures that gravitational-wave perturbation oscillate with a time-independent amplitude. In other words, the pure tensor modes are neutrally stable at the linear level on all scales.

\section{Non-linear gravitational-wave anisotropies}\label{sNlGWA}
The addition of large amplitude gravitational wave anisotropies can render the static universe unstable. This can be seen by examining Eq.~(\ref{frw}) for the isotropic and homogeneous background. Suppose that we add a simple anisotropy to the expansion, with isotropic 3-curvature. This is of the Bianchi I type and adds a new anisotropy energy density term for the shear, $\sigma^{2}=\Sigma^{2}/a^6$ with $\Sigma$ constant, to the Friedmann equation, where $a(t)$ is now the geometric mean scale factor and the Friedmann equation becomes
\begin{equation}
\left(\frac{\dot{a}}{a}\right)^2= \frac{\Sigma^{2}}{a^{6}}- \frac{\Psi}{a^{6}}+ \frac{\Gamma}{a^{4}}- \frac{1}{a^{2}}\,. \label{aniso}
\end{equation}
Clearly, if the anisotropy is sufficiently large, so that $\Sigma^{2}>\Psi$, then neither the static solution nor the exact solution with finite non-singular oscillations around it (\ref{eta})-(\ref{sol2}) any longer exist. This situation requires anisotropic perturbations to the FRW background that are far from linear.

We can envisage an irregular early state in which the ghost field, radiation, anisotropy and 3-curvature index, $K$, all vary in space. The effective Friedmann equation for the spatially-varying scale factor $a(\vec{x},t)$ would have the form
\begin{equation}
\left(\frac{\dot{a}}{a}\right)^2= \frac{\Sigma^{2}(\vec{x})}{a^{6}}- \frac{\Psi(\vec{x})}{a^{6}}+ \frac{\Gamma(\vec{x})}{a^{4}}- \frac{K(\vec{x})}{a^{2}}\,.
\label{aniso2}
\end{equation}
Regions where $\Sigma^{2}(\vec{x})>\Psi(\vec{x})$ would not oscillate through successive non-singular cycles; regions where $k(\vec{x})\leq0$ would bounce only once and then expand forever. Only those regions with low anisotropy and positive curvature would evolve through a succession of periodic cycles. If there was production of radiation entropy from cycle to cycle, in accord with the Second Law of thermodynamics, then this would be equivalent to an increase in the value of $\Gamma$ from cycle to cycle~\cite{BaDa} and a corresponding increase in the amplitude of the oscillations around the static universe in Eq.~(\ref{eta}), since they scale with $\Gamma$. They would approach the zero curvature expansion from below. If a positive cosmological constant were added to the right-hand side of Eq.~(\ref{aniso2}), then, no matter how small its magnitude, it would eventually come to control the dynamics if successive cycles grow in amplitude. Thereafter, the oscillations would cease and the expansion would asymptote towards an ever-expanding spatially-flat de Sitter state, as shown in~\cite{BaDa}. More complicated forms of anisotropy, including 3-curvature anisotropies, would display a similar instability because at early times they include also the simple anisotropic stress considered here.

\section{Discussion}
We have shown that in simple closed Friedmannian cosmologies containing black-body radiation and a ghost scalar field, there exist almost static solutions, which are oscillating about a finite radius. This is in sharp contrast to the instability shown by the Einstein-static cosmology, when conventional matter is used~\cite{E}. Our static universe also appears stable against all three type of linear inhomogeneous perturbations, in a way analogous to that seen in~\cite{Ha}-\cite{BEMT}. However, if large-amplitude homogeneous and anisotropic perturbations are introduced, the model becomes unstable (it no longer exists), with its past and future evolution moving towards a strong curvature singularity. In the absence of these large perturbations the static solutions with stable bounded oscillations could provide a viable past-eternal state for the Universe, although it would be subject to evolution if entropy was required to increase with time. Our analysis is of interest for studies of the very early universe incorporating theories of particle physics with new types of ghost matter field. It shows that it is theoretically possible for universes to get trapped in early-time evolutionary tracks that exhibit bounded oscillations about a non-singular static state. Similar behaviour might also arise in higher-order gravity theories~\cite{BH1}-\cite{K}, which introduce stresses that mimic the presence of ghost fields and would prevent some domains from participating in inflation.

\end{document}